\documentclass[12pt,a4paper]{article}
\usepackage{graphics}
\begin{document}
\baselineskip 0.25in
\title{\Huge{A Basic Introduction on Math-Link in Mathematica}}
\author{\Large {Santanu K. Maiti} \\ \\
         \Large {E-mail: {\em santanu.maiti@saha.ac.in}} \\ \\
          \Large {Theoretical Condensed Matter Physics Division} \\
           \Large {Saha Institute of Nuclear Physics} \\
            \Large {1/AF, Bidhannagar, Kolkata-700 064, India}}
\date{}
\maketitle
\newpage
\tableofcontents

\newpage
\begin{center}
\addcontentsline{toc}{section}{\bf {Preface}}
{\Large \bf Preface}
\end{center}

\noindent
Starting from the basic ideas of mathematica, we give a detailed 
description about the way of linking of external programs with mathematica 
through proper mathlink commands. This article may be quite helpful 
for the beginners to start with and write programs in mathematica. 

In the first part, we illustrate how to use a mathemtica notebook
and write a complete program in the notebook. Following with this, 
we also mention elaborately about the utility of the local and global
variables those are very essential for writing a program in mathematica. 
All the commands needed for doing different mathematical operations 
can be found with some proper examples in the mathematica book written 
by Stephen Wolfram~\cite{wolfram}.

In the rest of this article, we concentrate our study on the most
significant issue which is the process of linking of {\em external
programs} with mathematica, so-called the mathlink operation. By 
using proper mathlink commands one can run very tedious jobs efficiently
and the operations become extremely fast.

\newpage
\section{Introduction}

Mathematica is a robust software development for doing mathematics. It has
widespread applications in different fields and is often used for research, 
loading and analyzing data, giving technical presentations and seminars etc. 
Mathematica is extraordinary well-rounded. It is suitable for both numeric 
and symbolic work, and it has remarkable word-processing capabilities as 
well. Mathematicians can search for a working model, do intensive 
calculation, and write a dissertation on the project (including complex 
graphics) -- all from within mathematica. It is mathematica's complete 
consistency in design at every stage that gives it this multilevel 
capability and helps advanced usage evolve naturally.

Symbolic programming is the underlying technology that provides mathematica
this unmatched range of abilities. Just a single line sometimes makes a 
meaningful program in mathematica--the syntax, documents and methodology
used for input and output remaining as they are for immediate calculations. 
It supports every type of operation--be they data, functions, graphics, 
programs, or even complete documents--to be represented in a single, uniform 
way as a symbolic expression. This unification has many practical benefits 
to broadening the scope of applicability of each function. The raw algorithmic
power of mathematica is magnified and its utility extended.

\section{How to Start Mathematica ?}

We generally use mathematica through documents called {\em notebooks}.
To start a mathematica notebook in Unix we write `mathematica $\&$' 
from a command line and then press the `Enter' key from the key-board. 
A typical notebook consists of cells that may contain graphics, texts, 
programs or calculations. Now to exit from a mathematica notebook
we first go to the command `File' and then press `Quit' from the menu
bar of the notebook. Without using a notebook one can also use mathematica 
by typing the command `math' from a command line and all the jobs can 
also be done as well. To exit from mathematica for this particular case,
we should write either `Exit' or `Quit' and then press the `Enter' key. 
Thus one can run mathematica by using any one of the above two ways, but 
the most general way to do the interactive calculations in mathematica 
is the use of mathematica through notebook documents.

\subsection{How to Use a Mathematica Notebook ?}

In a notebook, a job is performed in a particular cell and for different 
jobs we use different cells. One can also use a single cell for all the 
operations, but it is quite easy if different operations are performed in
separate cells. A cell is automatically created when we begin to write
anything in the notebook. After writing proper operation/operations, it
is needed to run the jobs. For this purpose, we press the key `Shift' and
holding this key, we then press `Enter' from the key-board. The results
for the inputs are evaluated and they are available immediately underneath
in a separate cell, so-called the output cell.
                                                 
In mathematica we can do all kind of mathematical operations like numerical 
computation, algebric computation, matrix manipulation, different types of 
graphics etc., and all these things are clearly described by several examples 
in key mathematica book of {\em Wolfram Research}~\cite{wolfram}. So in
this article we shall not give any such example further. Now to do large
numerical computations, it is needed to write a complete program. For this
purpose, here we describe something about the way of writing a complete
program in a mathematica notebook.

\subsection{How to Write a Program in Mathematica ?}

In mathematica, we can write a program efficiently compared to any other
existing languages. As illustrative example, here we mention a very simple 
program which is: {\em the generation of a list of two random numbers and 
the creation of a $2$D plot from these numbers}.

\vskip 0.1in
\noindent
The program is:

\vskip 0.2in
\begin{center}
{\fbox{\parbox{5.65in}{\centering{
sample$[$times$_-]$$:=$Block$[\{$local variables$\}$,

numbers$=$Table$[\{$Random$[]$, Random$[]$$\}$, $\{i$, $1$, times$\}]$;

figure$=$ListPlot$[$numbers, PlotJoined$\rightarrow$True, 
AxesLabel$\rightarrow$$\{$xlabel, ylabel$\}$$]]$
}}}}
\end{center}

\vskip 0.12in
\noindent
This is the complete program for the generation of a list of two random 
numbers and the creation of a $2$D plot from these numbers. This program
is written in a single cell. After the end of this program we run it
by using the command `SHIFT' $+$ `ENTER', and then the mathematica does 
the proper operations and executes the result in an output cell.

Now to understand this program, it is necessary to describe the meaning
of the different commands used in this program. To start a program it is 
necessary to specify a name for the particular program. In this case, we 
specify `sample' as a program name, for the sake of simplicity. One can 
also use any other name in place of `sample', since this is a dummy name. 
If there is any running variable, like `times' (a dummy variable) in this 
particular case, then it has to be given within the bracket `$[~~]$'. After 
that the symbol `$_-$' is used, which indicates the variable as a functional 
variable. This is similar to define a functional variable, like $f[x]$ as 
$f[x_-]$ in mathematica. Now all the mathematical commands those are used 
for calculating the job are inserted within `Block$[\ldots ]$'. This is the 
central part of the program. This portion i.e., `Block$[\ldots ]$' is 
connected with `sample$[$times$_-]$' by the symbols `$:=$'. The symbol `$:$' 
has an important role, and therefore it has to be taken into account 
properly. Inside the `Block' the `local variables' for the program are 
\begin{figure}[ht]
{\centering \resizebox*{10.0cm}{6.5cm}{\includegraphics{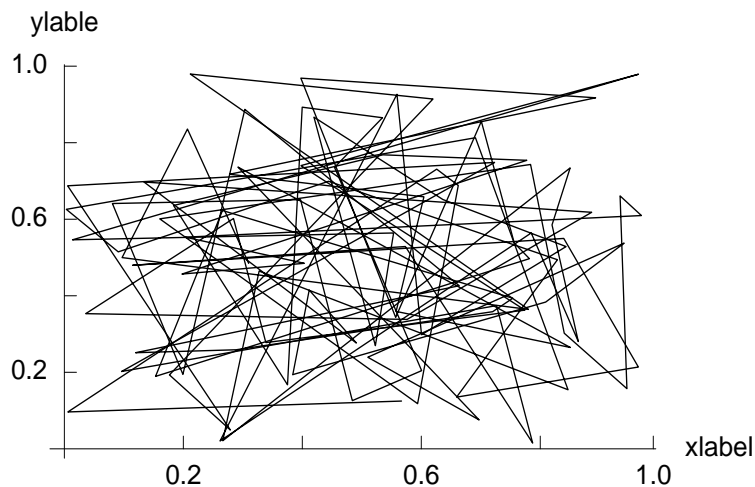}}\par}
\caption{A $2$D plot for a set of two random numbers.}
\label{random}
\end{figure}
declared within the bracket `$\{~~\}$'. There may also exist another type 
of variables called `global variables'. Later in this article, we will 
focus about these two different types of variables in detail. Now the rest 
part of the program differs from program to program depending on the 
nature of the particular operations. In this program, first we construct a 
list of two random numbers. In mathematica, a random number is generated 
simply by using the command `Random$[]$'. Therefore, a list of two 
such random numbers can be done very easily if we construct a table, 
which is performed by the command `Table' as given in the program. The 
integer $i$ runs from $1$ to `times', where the value of `times' can 
be put anything. So if we write `sample$[10]$', here `times $=10$', 
then $i$ runs from $1$ to $10$ and if we take `sample$[30]$', where
`times $=30$', then $i$ goes from $1$ to $30$. Now it becomes quite user
friendly if we mention different variable names for the different 
mathematical operations which are {\em not exactly} identical with 
any built in function available in mathematica like `Random', `Table', 
`Plot', etc. In this program we use the variable names `numbers' and 
`figure' for the two different operations. At the end of each mathematical 
operation, except the last operation which gives the final output of a 
program, we put the symbol `$;$'. This is also very crucial. Here we use 
the symbol `$;$' at the end of the second line only, but not in the last 
operation since this is the final output of this program. The command 
`ListPlot' plots the list of data points where the command 
`PlotJoined$\rightarrow$True' connects the lines between the data points. 
Finally, the command `AxesLabel' in this line is an option for the graphics 
functions to specify the labels in the axes. 

The output for this program is shown in Fig.~\ref{random}, which appears 
in a separate cell just below the input cell of the program. So now we 
can easily write and compile a program in mathematica. 

\subsubsection{Characterization of Local and Global Variables}

The local and global variables in mathematica play an important role, and 
therefore care should be taken about these two types of variables when we 
write a program in mathematica. We have already mentioned about the local
variables in the previous section that these variables are introduced only
inside the bracket `$\{~~\}$' at the beginning of the `Block$[~~]$'. In
such a case, the values of these parameters are only {\em defined within the 
cell} where we write a particular program. Outside this cell, they are
undefined and therefore, we may also use these same parameters for writing
other programs without any trouble.

On the other hand, the global variables are those which are not used within 
the bracket `$\{~~\}$' of a program. For such a case, these variables are 
assigned throughout the notebook for all cells. Thus if we declare any value
for a such parameter, then it will read this particular value whenever we
use it in any program. Accordingly, it may cause a difficulty if we use the 
same variable in another program by mistake. So we should take care about 
these two types of variables. To make it clear, here we illustrate the 
behavior of these two different kinds of variables by giving proper examples.

\vskip 0.2in
\begin{center}
{\fbox{\parbox{5.65in}{\centering{
sample$[$times$_-]$$:=$Block$[\{$$t=2.3,p=-1.5$$\}$,

numbers$=$Table$[\{$Random$[]$, Random$[]$$\}$, $\{i$, $1$, times$\}]$;

figure$=$ListPlot$[$numbers, PlotJoined$\rightarrow$True, 
AxesLabel$\rightarrow$$\{$xlabel, ylabel$\}$$]]$
}}}}
\end{center}

\vskip 0.12in
\noindent
Let us consider the above program which is written in a particular cell in
the mathematica notebook. In this program, we introduce two local variables 
$t=2.3$ and $p=-1.5$. Both these two variables are given inside the 
bracket `$\{~~\}$'. Now if we check the values outside the cell then the
output will be simply $t$ and $p$ for these two variables. So these are
the local variables and one can safely use these parameters again in other 
programs.

\vskip 0.2in
\begin{center}
{\fbox{\parbox{5.65in}{\centering{
sample$[$times$_-]$$:=$Block$[\{$$t=2.3,p=-1.5$$\}$,

$q=3.5$;

numbers$=$Table$[\{$Random$[]$, Random$[]$$\}$, $\{i$, $1$, times$\}]$;

figure$=$ListPlot$[$numbers, PlotJoined$\rightarrow$True, 
AxesLabel$\rightarrow$$\{$xlabel, ylabel$\}$$]]$
}}}}
\end{center}

\vskip 0.12in
\noindent
Now we refer to this program where we introduce an extra line for
another variable $q=3.5$ compared to the previous program of this section. 
Once we run this program, the value of $q$ will be assigned for any cell 
of the notebook. Therefore, in this case $q$ becomes the global variable, 
and if one uses it further in other program then the value of this parameter
$q$ will be assigned as $3.5$. Hence a mismatch will occur, and thus we 
should be very careful about these two different types of parameters.

\section{How to Link External Programs in Mathematica by Using Proper
         Math-Link Commands ?}

This section describes the most significant issue of this article which 
deals with the way of linking of an external program with mathematica 
through proper mathlink commands. The mechanism for the linking of 
external program written in C with mathematica has already been 
established~\cite{maeder}. But this will not work if one tries to link 
an external program written in other languages like F77, F90, F95, etc.,
with mathematica. This motivates us to find a way of linking an external
program written either in any one of these later languages (F77, F90, F95)
with mathematica. Here we illustrate it for the FORTRAN-90 source 
files~\cite{smith,mart} only, but this mechanism will also work 
significantly for the other Fortran source files as well.

\subsection{Mathlink for XL Fortran-90 Source Files}

In order to understand the basic mechanism for linking an external program
with mathematica, let us begin by giving a simple example. Here we set the
program as follows:

\begin{enumerate}
\item Construct two square matrices in mathematica.
\item Take the product of these two matrices by using an external program
written in F90.
\item Calculate the eigenvalues of the product matrix in mathematica.
\end{enumerate}

\vskip 0.12in
\noindent
The whole operations can be pictorially represented as,
\begin{figure}[ht]
{\centering \resizebox*{10.0cm}{3.5cm}{\includegraphics{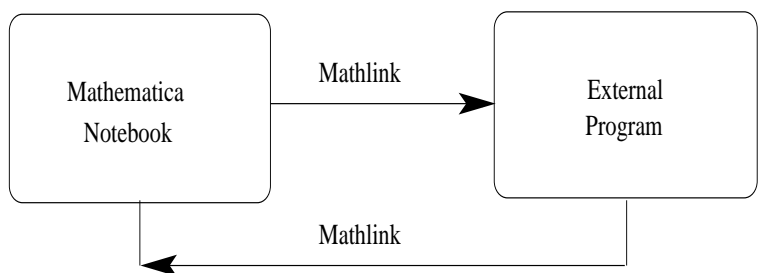}}\par}
\caption{Schematic representation of mathlink operations.}
\label{diagram}
\end{figure}

\vskip 0.12in
\noindent
The operations $1$ and $3$ are performed in mathematica, while the operation
$2$ is evaluated by the external program. The transformations of the datas
from the mathematica notebook to the external program are done by using
some proper commands, so-called mathlink operation. To complete this 
particular job (operations $1$-$3$), we need two programs. One is written
in mathematica for the operations $1$ and $3$, while the other program is
written in F90 for the operation $2$. Now we describe all these steps one
by one. Let us first concentrate on the external program, given below, 
where the multiplication of the two square matrices (operation $2$) is 
performed. The first line of the program corresponds to the command line 
where the symbol `!'
is used to make a statement as a command statement. The next line provides
a specific name of the program which is described by the command `program
multiplication'. This actually starts the program, and accordingly, the
program is ended by the command `end program multiplication'. In F90,
we can allocate and deallocate array variables in the programs which help
us a lot to save memory and are very essential to run many jobs 
simultaneously. Here we use three array variables `a, b and c' for the 
three different matrices whose dimensions are allocated by the order of 
the matrix `n'. Finally, the product of the two matrices `a' and `b' is
determined by the command `matmul(a,b)' and the datas are stored in the
matrix `c'. This is the full program for the matrix multiplication of any
two square matrices of order `n'.

\vskip 0.2in
\begin{center}
{\fbox{\parbox{4.75in}{\centering{
! A program for matrix multiplication of two square matrices

program multiplication

implicit double precision (a-h,o-z)

double precision, allocatable :: a(:,:),b(:,:),c(:,:)

read * , n !(the order of the matrix)

allocate (a(n,n),b(n,n),c(n,n))

read * , ((a(i,j),j=1,n),i=1,n) ; read * , ((b(i,j),j=1,n),i=1,n)

!    Calculation of matrix multiplication :

c=matmul(a,b)

print'(1(1x,f10.6))',((c(i,j),j=1,n),i=1,n)

end program multiplication
}}}}
\end{center}

\subsubsection{How to Compile and Optimize XL Fortran-90 Source Files ?}

After writing a program, first we need to compile it to check whether 
there is any syntax error or not to proceed for further operations. 
Several commands are accessible for the compilation and optimization of 
a program. The commands generally used to compile a F$90$ source file 
are: xlf$90$, xlf$90_-$r, xlf$90_-$r$7$, etc. Thus we can use anyone of 
these to compile this program, but different commands optimize a program 
in different ways which solely depends on the nature of the particular 
program. The simplest way for the compilation of a program is,

\vskip 0.2in
\begin{center}
{\fbox{\parbox{1.75in}{\centering{
xlf90 filename.f 
}}}}
\end{center}

\vskip 0.12in
\noindent
With this operation, an `executable file' named `a.out' is created,
by default, in the present working directory (pwd). But if one uses several 
programs simultaneously then it would be much better to specify different 
names of different `executable files' for separate programs. To do this 
we use the prescription,

\vskip 0.2in
\begin{center}
{\fbox{\parbox{2.55in}{\centering{
xlf90 filename.f -o filename
}}}}
\end{center}

\vskip 0.12in
\noindent
Under this process, the `executive file' named as `filename' is created. 
Thus we can create proper `executive files' for different jobs and all
the jobs can be performed simultaneously without any difficulty.
 
For our illustrative purposes, below we mention some other optimization 
techniques for the Fortran source files.

\begin{itemize}
\item -o : Optimizes code generated by the compiler.

\item -o0 : Performs no optimizations. (It is the same as -qnoopt.)

\item -o2 : Optimizes code (this is the same as -O).

\item -o3 : Performs the -O level optimizations and performs additional
      optimizations that are memory or compile time intensive.

\item -o4 : Aggressively optimizes the source program, trading off
      additional compile time for potential improvements in the
      generated code.  This option implies the following options:
     -qarch=auto -qtune=auto -qcache=auto -qhot -qipa. 

\item -o5 : Same as -O4, but also implies the -qipa=level=2 option.
\end{itemize}

\vskip 0.12in
\noindent
From these operations, we can make some flavors about the compilation 
and optimization technique for a Fortran source file. For a detailed 
description of each operation, we refer to the XL Fortran User's 
Guide~\cite{ibm}.

\subsubsection{Link of XL Fortran-90 Program with Mathematica}

This is the heart of this article. Below we set the mathematica program
for the operations $1$ and $3$, incorporating the operation $2$ by using
the proper mathlink commands, and illustrate all the steps properly.

Let us suppose the external program, for the operation $2$, is 
written in the directory `/allibmusers/santanu/files/test'. Generally we 
are habituated to see the working directory as `/user/santanu/...' or 
`/home/santanu/...' or `/allusers/santanu/...', etc. So it can be anything 
like these. Thus knowing the directory where the external program is written,
we enter into that particular directory and compile the external program
properly to create an `executive file' for further operations. For this
particular case, we create the `executive file' named as `mat' which is
used in the $13$-th line of the following mathematica program. Now the 
external program is ready for the operation, and we enter into the  
directory where we will run the job in the mathematica notebook for the
operations $1$ and $3$.

Sitting in the directory where the mathematica notebook is open, we 
need to connect the proper directory where the `executive file' for 
the external program exists. The name of the pwd can be checked 
directly from the mathematica notebook by using the command `Directory[]'. 
Suppose the pwd is `/allibmusers/santanu/math'. Now If this pwd is 
different from the directory where the file `mat' exists, then we make
a link to that particular directory through the command `SetDirectory'. 
Below we give an example to connect the directory 
`/allibmusers/santanu/files/test', where the file `mat' exists.

\vskip 0.2in
\begin{center}
{\fbox{\parbox{4in}{\centering{
SetDirectory[``/allibmusers/santanu/files/test"]
}}}}
\end{center}

\vskip 0.12in
\noindent
For this operation, the total path must be used within the double 
quotes `` ". Using the command `ResetDirectory[]', we can come back to 
the initial directory. Thus we can connect and disconnect any directory
with the pwd from the mathematica notebook, and able to link external 
programs with mathematica very easily. 

\vskip 0.2in
\begin{center}
{\fbox{\parbox{5.5in}{\centering{
sample$[$ns$_-]$$:=$Block$[\{$$t=0,s=0$,vacuum1$=\{\}$,vacuum2$=\{\}$$\}$,

SetDirectory[``/allibmusers/santanu/files/test"];

Do[Do[a1 = If[i == j, t, 1.213]; \\
    a2 = AppendTo[vacuum1, a1], $\{$j, 1, ns$\}$], $\{$i, 1, ns$\}$]; \\
mat1 = Partition[a2, ns];

Do[Do[a3 = If[k == l, s, 2.079]; \\
      a4 = AppendTo[vacuum2, a3], $\{$l, 1, ns$\}$], $\{$k, 1, ns$\}$]; \\
mat2 = Partition[a4, ns];

mat3 = Partition[Flatten[$\{\{$mat1$\}$, $\{$mat2$\}\}$], ns]; \\
matrixorder = $\{$ns$\}$; \\
output = Insert[mat3, matrixorder, 1]; \\
Export[``mat3.dat", output];

matrix = Partition[
   Flatten[ReadList[``$!$mat$<$mat3.dat", Number, \\
             RecordLists$\rightarrow$ True]], ns]; \\
  results = Eigenvalues[matrix]$]$
}}}}
\end{center}

\vskip 0.12in
\noindent
In the above program, the variables `t' and `s' are the local variables, 
and we have already discussed about these variables in the previous section. 
`vacuum1=$\{\}$' and `vacuum2=$\{\}$' are the two empty lists where the 
datas are stored for each operation of the two `DO' loops given in the 
program to make the lists `a2' and `a4' respectively. The `Partition' 
command makes the partition of a list. The parameter `ns' gives the order 
of the two square matrices. By using the command `Export' we send the file 
`mat3.dat' which is treated as the input file for the external program 
kept in the directory `/allibmusers/santanu/files/test'. To perform the 
matrix multiplication by using the external program and get back the 
product matrix in the mathematica notebook we use the operation: 
ReadList[``!mat$<$mat3.dat", Number, RecordLists$\rightarrow$True].
Here the command `ReadLeast' is used to read the objects from a file
and the commands `Number' and `RecordLists$\rightarrow$True' are the
options of the command `ReadList'. Finally, the eigenvalues of the
matrix in the mathematica notebook are determined by using the 
command `Eigenvalues'.

\subsubsection{Link of other XL Fortran Programs with Mathematica}

Now we can also use the mathlink operations for other programs written 
either in F$77$ or F$95$ by the above mechanisms. For these programs, we 
should use proper commands for the compilations and optimizations. As 
representative example, here we mention some of the commands for the 
compilation of these XL Fortran source files those are: xlf, f$77$, 
fort$77$, xlf$_-$r, xlf$_-$r$7$, xlf$95$, xlf$95_-$r and xlf$95_-$r$7$.

So now we can able to use mathlink commands for any type of Fortran program.

\noindent
\addcontentsline{toc}{section}{\bf {Concluding Remarks}}
\begin{flushleft}
{\Large \bf {Concluding Remarks}}
\end{flushleft}
\vskip 0.1in
\noindent
In conclusion, we have explored some significant issues of mathematica
starting from its basic level i.e., how to start mathematica, open a 
mathematica notebook, write a program in mathematica, etc., and finally,
illustrated the basic mechanism for the linking of external programs with 
mathematica notebook. This mathlink operation is the heart of this article,
and it is extremely crucial for doing large numerical computations. 
In this article, we have concentrated the mathlink operations mainly for 
the XL Fortran 90 source files. But these operations can also be used for
any other Fortran source files. Following with this, we have also 
illustrated very briefly about the optimization techniques for the Fortran
source files which may help us to run very complicated jobs quite 
efficiently.

\noindent
\addcontentsline{toc}{section}{\bf {Acknowledgment}}
\begin{flushleft}
{\Large \bf {Acknowledgment}}
\end{flushleft}
\vskip 0.1in
\noindent
I acknowledge with deep sense of gratitude the illuminating comments and
suggestions I have received from Prof. Sachindra Nath Karmakar during the 
preparation of this article.

\addcontentsline{toc}{section}{\bf {References}}


\begin{thebibliography}{99}

\bibitem{wolfram} Stephen Wolfram. {\em Mathematica-5.0}.
\bibitem{maeder} Roman E. Maeder. {\em About Parallel Computing Toolkit}.
         A Wolfram Research Application Package.

\bibitem{smith} I. M. Smith. {\em Programming in Fortran 90 : A First Course
         for Engineers and Scientists}. University of Manchester, UK.
\bibitem{mart} Martin Counihan. {\em Fortran 90}. University of Southampton.
\bibitem{ibm} IBM. {\em XL Fortran for AIX : User's Guide}.

\end{thebibliography}
\end{document}